# Stars of the Lower Part of the Main Sequence with Discovered Exoplanets and Candidates.
# Periods of Rotations or Revolutions?

*Aleksey A. Shlyapnikov*
*ass@craocrimea.ru*

**Abstract**
The purpose of this work is to supplement the "Stars with solar-type activity" catalog with information about confirmed exoplanets and exoplanet candidates. To do this, cross-identification of the catalog stars with data from the NASA exoplanet archive was carried out. This article presents the distribution of the number of suspected stars with exoplanets by brightness, spectral type, amplitude of variability, as well as other statistical analysis data. Particular attention has been paid to the comparison of the periods of rotation of stars and the orbital periods of revolution of exoplanets around them. Analysis of the data suggests the need to change the previously determined types of variability in some stars.

## 1. Introduction

The results of a statistical analysis of data on stars with discovered exoplanets were previously published in an article by the author (Shlyapnikov 2022). The objects are included in the new "Catalog of Stars with Solar Type Activity", hereinafter referred to as CSSTA[1]). The catalog contains data on 314618 objects from the lower part of the Main Sequence. It provides information about coordinates, identification, types of variability, magnitudes, amplitudes of variability, spectral types, the presence of radiation in the X-ray, ultraviolet and infrared ranges, and other information about the physical parameters of stars. The 2020 version of the CSSTA catalog is described in more detail in the monograph (Gershberg et al., 2020).

The presence of X-ray and radio emissions, if any, should contribute to the understanding of the possibility of the existence of "life" in the habitable zone. Flare activity, if present, imposes certain restrictions on the development of life. The periods of revolution of exoplanets around stars, and the associated changes in brightness, albeit insignificant, should be taken into account when analyzing its own rotation and the period of possible cyclic activity.

By early February 2023, according to the NASA Exoplanet Archive[2] the existence of 5250 planets was confirmed, as discovered by various 11 methods, and 291 found by the TESS observatory[3]. The discovery of another 6153 exoplanets according to TESS observations needs to be confirmed.

Information about the study of exoplanets at the Crimean Astrophysical

---

[1] CSSTA – http://craocrimea.ru/~aas/CATALOGUEs/CSAST/CSAST.html
[2] NASA exoplanet archive – https://exoplanetarchive.ipac.caltech.edu
[3] TESS Exoplanet Mission – https://www.nasa.gov/tess-transiting-exoplanet-survey-satellite

Observatory was described in (Moskvin et al., 2018) and posted on a specially created website[4].

## 2. Statistics for stars with exoplanet candidates

To conduct a statistical analysis of the characteristics of stars with suspected exoplanets, we performed cross-identification of CSSTA objects with information from the NASA exoplanet archive for candidates discovered from observations by the TESS observatory. All observational data were obtained by the transit method, that is, by analyzing the change in the brightness of a star as an exoplanet passes through its disk.

The 6153 exoplanet candidates belong to 5941 stars. We note that some stars contain more than one candidate; some of the candidates, after confirmation, were transferred to the database of planetary systems.

The distribution of the number of stars with exoplanet candidates by magnitude is shown in Figure 1. As can be seen on it, the distribution maximum is in the range from 11 to 12 magnitudes. This is due to the use of short-focus lenses and low exposures in the sky survey performed by TESS in order to search for exoplanet transits. Figure 2 illustrates the distribution of variability amplitudes for stars with suspected exoplanets. The maximum in the distribution of the number of stars from the brightness variation amplitude is within $0^m.001$. Reducing the number of objects in an area with an amplitude of about $0^m.006$ with a subsequent increase may indicate a transition of the observed brightness changes from those associated with the transit of an exoplanet to those associated with the rotation of the star.

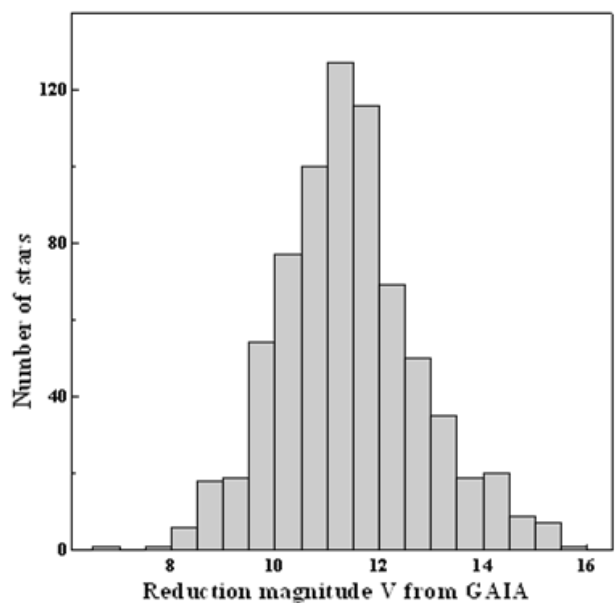 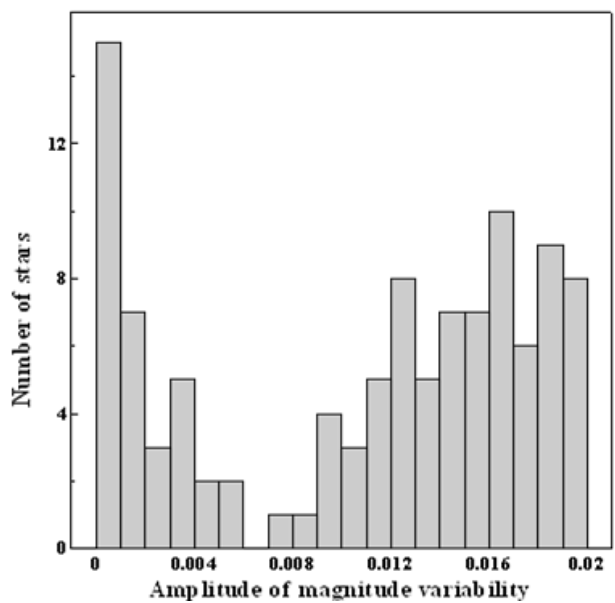

Fig. 1. The distribution of the number of stars with exoplanet candidates by brightness included in the CSSTA.

Fig. 2. The distribution of the number of stars over the amplitudes of the detected variability.

---

[4] CrAVO exoplanet – https://sites.google.com/view/cravo-exop-psa

Figure 3 shows the distribution of the number of considered stars and CSSTA objects depending on their spectral types. The maximum in the distribution is in the G9–K0 range. We note that this sample of objects identified by the CSSTA does not contradict the spectral classification of all stars with exoplanets described in (Gorbachev et al. 2019, Shlyapnikov 2022). The distribution maxima in these works are also close to the spectral type K.

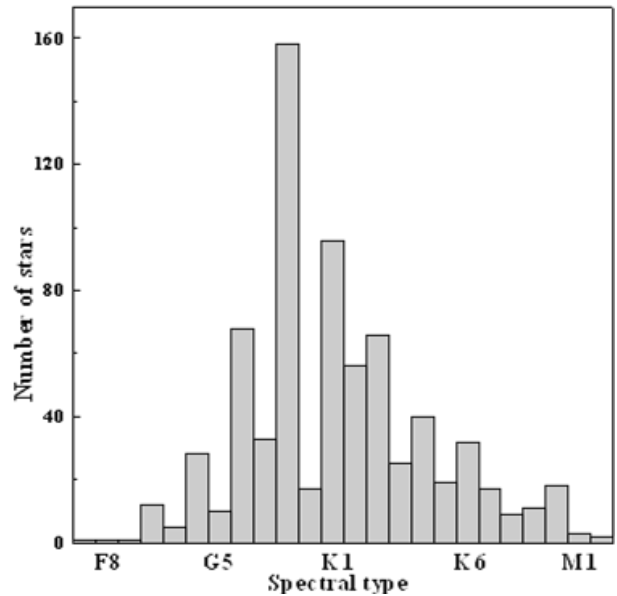

Fig. 3. The spectral type distribution of the number of CSSTA stars with exoplanet candidates.

**3. Periods of rotations or revolutions?**

An analysis of periodic variations in the brightness of the stars under consideration is of particular interest. Figures 4 and 5 show a histogram of the distribution of variability periods for stars with exoplanet candidates according to VSX data[5] (Watson et al., 2006) and a histogram of the distribution of orbital periods of planets around these stars according to the NASA Exoplanet Archive. The maximum in the distribution according to the VSX data is 3.5 days (Fig. 4), how and the second largest maximum in the distribution of candidate circulation periods (Fig. 5). At the same time, it should be noted that a small part of the considered stars with unconfirmed exoplanets according to the SIMBAD databases[6] (Wenger et al., 2000) and VSX have variability due to rotational modulation of the spotted surface. These are stars like BY Dra[7] and RotV∗[8].

---

[5] The International Variable Star Index (VSX) - https://www.aavso.org/vsx/index.php
[6] Astronomical database SIMBAD - http://simbad.cds.unistra.fr/simbad/
[7] Variable of the BY Dra type.
[8] A star whose variability is due to the rotational modulation of the spotted surface.

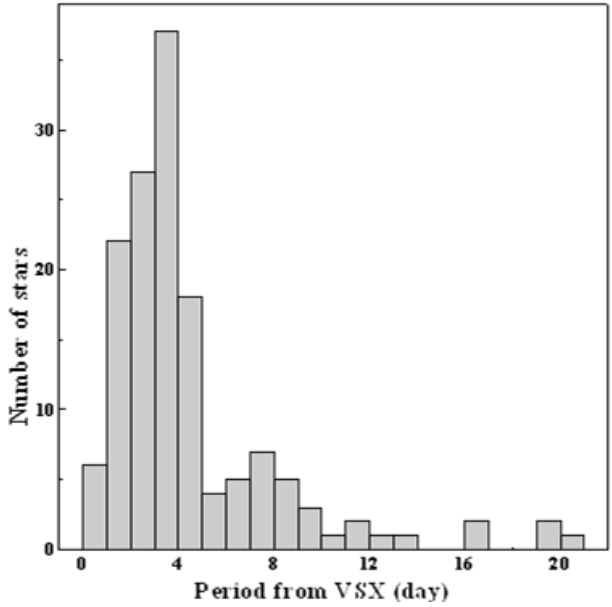 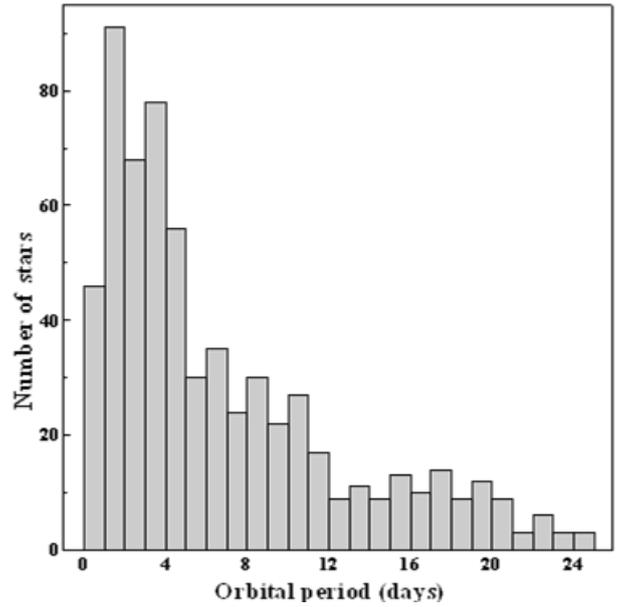

Fig. 4. The period distribution of stars with planets according to VSX data.

Fig. 5. The orbital periods of exoplanets around stars.

The detected brightness variability of stars in the lower part of the Main Sequence is most characteristically demonstrated not by the features of rotational modulation, but by the presence of exoplanets (Figs. 6 and 7).

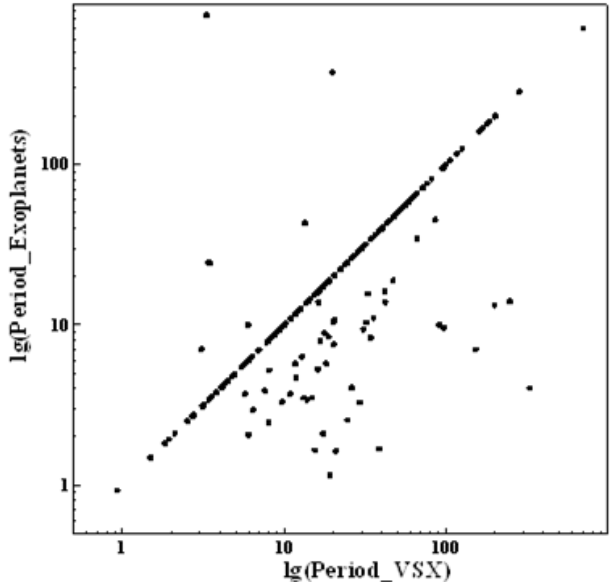 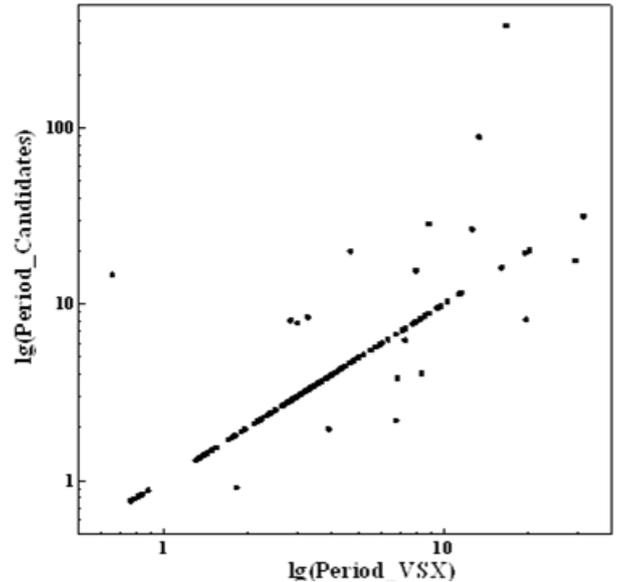

Fig. 6. The dependence of the orbital periods of confirmed exoplanets on the periods of change in the brightness of their host stars.

Fig. 7. The dependence of the orbital periods of exoplanet candidates on the periods of change in the brightness of the host stars.

Figure 6 is based on the data presented in (Shlyapnikov 2022) for stars with confirmed exoplanets; 25 objects considered in the article have the BY Dra variability type according to the SIMBAD database, and 467 have RotV*, i.e., and in the first and second cases, their variability is due to the rotational modulation of a spotted surface, and not to the presence of exoplanets. Since the SIMBAD astronomical database is built on the basis of including information from publications, the type of stellar variability was determined from real observations. Figure 6 illustrates that more than half of the objects for which periods have been determined as rotation periods are probably exoplanet orbital periods around their stars (virtually linear dependence of periods in Fig. 6). A similar picture is observed when considering the ratio of periods for stars in which planetary candidates have not yet been confirmed (Fig. 7). Obviously, the periods that do not fit into a linear relationship in both figures are a reflection of real changes in the physical conditions on the stars, which are observed as a change in their brightness. One possible reason for the coincidence of the periods of rotation of stars and the circulation of exoplanets can be their synchronization. In this case, the presence of planets should have a direct impact on the manifestation of stellar activity.

## 4. Conclusions

After cross-identification between the CSSTA catalog data and the NASA Exoplanet Candidate Archive, the catalog was supplemented with information for 689 stars. An analysis of the distribution of the number of studied objects from the magnitude showed that the maximum is in the range from $11^m$ up to $12^m$.

Consideration of the distribution of brightness variability amplitudes of stars in which exoplanets are suspected showed that its maximum is within $0^m.001$. The increase in the amplitude of variability after $0^m.006$ possibly indicates a transition of the observed brightness changes from those associated with the transit of an exoplanet to those associated with the rotation of a star.

An analysis of the distribution of the number of stars under consideration depending on their spectral types confirmed that its maximum is located in a region close to the spectral type K. This is consistent with previous studies.

A study of periodic brightness changes found in stars with confirmed and suspected exoplanets showed that most of the periods found for rotationally modulated variables correspond to the periods of rotation of the planets around these stars. This analysis of the data suggests the need to revise the previously determined types of variability for some stars.

**Acknowledgments**

This research has made use of the SIMBAD database, operated at CDS, Strasbourg, France; the International Variable Star Index (VSX) database, operated at AAVSO, Cambridge, Massachusetts, United States; the NASA Exoplanet Archive, which is operated by the California Institute of Technology, under contract with the National Aeronautics and Space Administration under the Exoplanet Exploration Program.